\documentstyle[amssymb,prl,epsf,aps,multicol]{revtex}

\begin{document}

\sloppy

\draft

\title{Modified critical correlations close to modulated and rough surfaces}
\author{\\Andreas Hanke and Mehran Kardar}
\address{Department of Physics, Massachusetts Institute of Technology,\\
Cambridge, Massachusetts 02139}
\date{\today}
\maketitle
\begin{abstract}
Correlation functions are sensitive to the presence of a boundary.
Surface modulations give rise to modified near surface correlations,
which can be measured by scattering probes. To determine these
correlations, we develop a perturbative calculation in deformations
in height from a flat surface. The results, combined with a 
renormalization group around four dimensions, are also used to 
predict critical behavior near a self-affinely rough surface. 
We find that a large enough roughness exponent can modify surface 
critical behavior.
\end{abstract}

\begin{multicols}{2} \narrowtext
Bulk properties, such as magnetization, as well as correlation 
functions are modified on approaching a surface. In particular, 
critical behavior near surfaces or defects, which is quite 
different from the bulk, has been extensively studied theoretically 
\cite{Bin83,Die86,DW84,AH2000}, by experiments \cite{MDPJ90,Dosch92,exp}, 
and in simulations \cite{simu}. Along with this development, 
the method of grazing incidence of x-rays and neutrons \cite{DW84}
has become a standard tool of probing critical behavior near surfaces 
and interfaces \cite{MDPJ90,Dosch92,exp}.

Most theoretical studies have been restricted to flat surfaces. 
However, real surfaces mostly deviate from this idealized picture.
Possible deviations can be divided into two classes:
(i) advanced experimental methods, e.g., lithographic preparation,
allow one to endow surfaces with specific, regular geometrical
patterns down to the nanometer scale, with important applications 
in technology and material science \cite{Whi97}; 
(ii) surfaces or interfaces can be naturally rough, e.g., 
due to growth, fracture, or erosion. Most common are self-affinely 
rough surfaces, in which the root mean square height fluctuations on 
a length scale $L$ grow as $L^{\zeta}$, where $\zeta < 1$ is the 
so-called roughness exponent. Self-affine scaling is predicted by 
many numerical and analytical models of surface growth \cite{Kar96}, 
and is also observed in a number of experiments \cite{KP95}.

In this work we show that the shape of the surface has a distinct 
influence on the properties of an adjacent medium with long-range 
correlations. This is demonstrated for two-point correlation functions 
near a critical point of the medium, for both cases (i) and (ii) 
outlined above. 
The diffuse scattering of x-rays and neutrons at grazing incidence 
due to such correlations appears in addition to what would have 
been observed if the surface was separating two homogeneous media 
\cite{DH95}. The modified correlations may thus provide an additional 
and indirect means of characterizing the surface profile. This may 
be of value when other techniques are not possible, as in the case 
of the interior surface of a glass, or an internal crack, whereas 
scattering from a critical fluid or binary alloy coating the surface 
may be feasible.

In order to study the effects of the surface shape, 
we develop a perturbative expansion of
two-point correlation functions in the deformations of the 
height profile. Initially for a Gaussian field,
the calculations are carried out to second order.
Already at the first order, the two-point correlation functions
track the profile from the substrate, with a modulation that
decreases with the distance of the two points from the surface. 
This leads to explicit predictions for the structure factor, 
as a function of the lateral wave vector transfer, for a 
modulated surface [see Eqs.\,(\ref{s1}) and (\ref{s2}) below].
For example, for a sine modulation 
with wavelength $2 \pi / k$ along one direction, the incident 
wave vector is scattered by $k$ in that direction, with 
amplitude proportional to the modulation in height.

For self-affinely rough surfaces, second order calculations 
are necessary, as the first order results vanish on average.
For a massless Gaussian field we find the expected result that
self-affine roughness leads to subleading corrections to the decay of 
two-point correlation functions, which at a scale $r$ are smaller
by a factor of $r^{-2(1-\zeta)}$ than the leading contribution 
coming from a flat surface.
Typical critical systems, however, are described 
by a non-Gaussian (interacting) field theory. In this case, the 
correlations are calculated perturbatively in the strength of the 
interaction, and the results interpreted with the aid of the
renormalization group (RG) in $4 - \varepsilon$ dimensions
[see Eqs.\,(\ref{perp}) - (\ref{etaparalleltilde}) below].
We find that the subleading corrections now fall off with 
a slower power as compared to the Gaussian case
and, surprisingly, for a sufficiently large $\zeta$ even
{\em dominate\/}, giving rise to novel surface critical behavior.

Fluctuations in the critical system located above and bounded by
the surface will be described by an $n$-component field 
$\Phi(\underline{r}) = 
[\Phi_1(\underline{r}), \ldots , \Phi_n(\underline{r})]$,
where $\underline{r}$ is assumed to be $d$ dimensional. For 
example, $n = 1$ can represent an Ising magnet or binary alloy, 
while $n = 2$ may describe a superfluid. The fluctuations are
described by the statistical Boltzmann weight
$e^{- \beta {\cal H}}$, with
\begin{equation} \label{hamiltonian}
\beta {\cal H}\{ \Phi \} \, = \,   
\int d^d r \, \left\{ \frac{1}{2} (\nabla \Phi)^2 \, + \,
\frac{\tau_0}{2} \, \Phi^2 
\, + \, \frac{u_0}{4!} \, (\Phi^2)^2 \right\}
,
\end{equation}
where $u_0$ is the strength of the interaction, set to zero
in the Gaussian theory, and $\tau_0 \sim T - T_c$.
The above expression must be supplemented 
by a boundary condition on the surface. We choose the Dirichlet 
boundary condition $\Phi = 0$, which represents the so-called 
ordinary surface universality class, appropriate to magnets, 
binary alloys, and for $^4$He near the normal to 
superfluid transition point \cite{Bin83,Die86,DLBD97}.

\begin{figure}[t]
\begin{picture}(8,5)
\put(-0.5,0.7){
\setlength{\epsfysize}{4.3cm}
\epsfbox{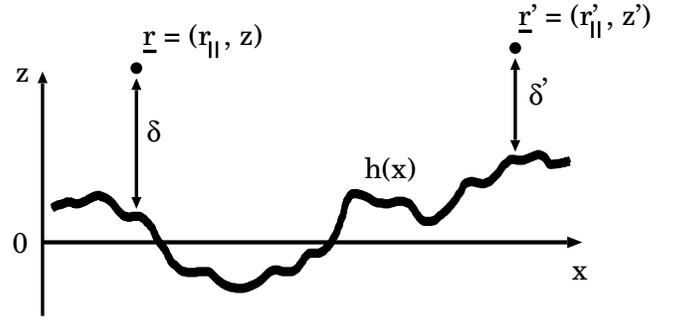}}
\end{picture}
\caption{Position vectors 
$\underline{r} = ({\bf r}_{\parallel}, z)$ and 
$\underline{r}' = ({\bf r}_{\parallel}', z')$ of the   
two-point correlation function in the critical system located 
above and bounded by a deformed surface. The surface profile is
described by the height function $h({\bf x})$ and the vertical
distances of $\underline{r}$ and $\underline{r}'$ from the surface are 
given by $\delta = z - h({\bf r}_{\parallel})$ and  
$\delta' = z' - h({\bf r}_{\parallel}')$, respectively.}
\label{fig1}
\end{figure}

In the absence of overhangs and inlets, the surface profile can be
described by a single-valued 
height function $h({\bf x})$, where ${\bf x}$
spans a $D = d - 1$ dimensional base plane (see Fig.\,\ref{fig1}).
The Gaussian correlation 
$\langle \Phi_i(\underline{r}) \Phi_j(\underline{r}') \rangle \, = \, 
\delta_{ij} \, G(\underline{r} ; \underline{r}')$, according to  
Eq.\,(\ref{hamiltonian}) with $u_0 = 0$,
can be calculated using functional integral methods 
\cite{LK91,GK97} and is given by
\begin{eqnarray}
G(\underline{r}; \underline{r}') \, & = & \, 
G_b(\underline{r}; \underline{r}') \, -
\int d^D x \int d^D y  \label{gauss2} \\
& & \times \, G_b(\underline{r} ; {\bf x}, h({\bf x})) \,
M({\bf x}, {\bf y}) \, G_b(\underline{r}' ; {\bf y}, h({\bf y})) 
\, \, . \nonumber
\end{eqnarray}
Note that we denote $d$ dimensional vectors with underlined 
letters, and $D$ dimensional vectors with boldface letters.
Position vectors $\underline{r}$ are thus decomposed according to 
$\underline{r} = ({\bf r}_{\parallel}, z)$, where 
${\bf r}_{\parallel}$ comprises the $D = d - 1$ components 
parallel to the surface and $z$ is the distance from the 
base plane;
$G_b$ is the bulk correlation function of the Gaussian 
theory, and the kernel $M({\bf x}, {\bf y})$ satisfies
\begin{equation} \label{inverse}
\int d^D y \, M({\bf x}, {\bf y}) \, 
G_b({\bf y}, h({\bf y}); {\bf y}', h({\bf y}')) = 
\delta^D({\bf x} - {\bf y}') \, .
\end{equation}
While the above results are generally valid,
we focus on the behavior of the 
correlation functions at the bulk critical point, i.e., 
for $T = T_c$, where correlations are strongest.

The solution for $G$ can be expanded in a series 
$G_0 + G_1 + G_2 + \ldots$ in powers of $h({\bf x})$. 
The lowest order result,
$G_0(\underline{r} ; \underline{r}') = 
G_b({\bf r}_{\parallel} , z ; {\bf r}_{\parallel}', z') -
G_b({\bf r}_{\parallel} , z ; {\bf r}_{\parallel}', - z')$,
corresponds to a flat surface.
The bulk correlation function
$G_b(\underline{r} ; \underline{r}')$ decays as $r^{-(d-2+\eta)}$
for large separations $r = |\underline{r} - \underline{r}'|$, 
where the bulk critical exponent $\eta$ is given by
$\eta = 0$ in the Gaussian theory.
In contrast, if both points remain
close to the surface, $G_0(\underline{r} ; \underline{r}')$ decays as 
$r^{-(d-2+\eta_{\parallel})}$, where $\eta_{\parallel}$
is a surface critical exponent given by 
$\eta_{\parallel} = 2$ in the Gaussian theory 
\cite{Bin83,Die86}.

The first order result is given by
\begin{equation} \label{h1}
G_1(\underline{r} ; \underline{r}') = -
\int d^D x \, \Delta({\bf r}_{\parallel} - {\bf x} , z)
\, h({\bf x})
\Delta({\bf r}_{\parallel}' - {\bf x} , z') \, , \nonumber
\end{equation}
where
$\Delta({\bf x} , z) \, = \, 
\int \frac{d^D p}{(2 \pi)^D} \,
e^{i {\bf p} \cdot {\bf x}} \, e^{- p z}$,
with $p = |{\bf p}|$,
has the form of a representation of the 
delta function $\delta^D({\bf x})$, i.e.,
$\int d^D x \, \Delta({\bf x}, z) = 1$ and
$\lim_{z \to 0} \, \Delta({\bf x}, z) = \delta^D({\bf x})$.
The first order result $G_1$ tracks the profile $h({\bf x})$
of the surface. 

%%%%%%%%%%%%%%%%%%%%%%%% Fig.: figure1.ps %%%%%%%%%%%%%%%%%%%%%%%%%
%
\unitlength1cm
%
%%%%%%%%%%%%%%%%%%%%%%%%%%%%%%%%%%%%%%%%%%%%%%%%%%%%%%%%%%%%%%%%%%%%

\smallskip

\noindent
For example, for 
$r = |\underline{r} - \underline{r}'| \to \infty$
with $z$ and $z'$ fixed, the results for 
$G_0$ and $G_1$ imply the behavior
\begin{equation} \label{leading}
G(\underline{r} ; \underline{r}') \sim  
[ 1 - A(\underline{r}) - A(\underline{r}') ] \, 
r^{- (d - 2 + \eta_{\parallel})}
,
\end{equation}
up to terms of order $(h/z)^2$ and $(h/z')^2$. 
Thus, the leading power law is the same as for a flat surface,
but the amplitude is modulated by the surface deformations in the
vicinity of ${\bf r}_{\parallel}$ and ${\bf r}_{\parallel}'$ 
by \cite{rem}
\begin{equation} \label{A1}
A(\underline{r}) \, = \,
\frac{\eta_{\parallel} - \eta}{2} \,
\int d^D x \, \frac{h({\bf x})}{z} \,
\Delta( {\bf x} - {\bf r}_{\parallel}, z) \, \, .
\end{equation}

Already the results at first order indirectly 
characterize the surface in scattering experiments. To 
demonstrate this, we introduce the Fourier transform
of the height profile as
$h({\bf x}) = \int \frac{d^D k}{(2 \pi)^D} \,
e^{i {\bf k} \cdot {\bf x}} \, \hat{h}({\bf k})$,
with $\hat{h}(-{\bf k}) = \hat{h}({\bf k})^{\ast}$, 
and accordingly the lateral structure factor 
$S({\bf p}, z; {\bf p}',z')$
corresponding to $G(\underline{r}, \underline{r}')$
(where the parallel component 
${\bf r}_{\parallel}$ is transformed to the lateral 
wave vector ${\bf p}$). 
The first order result in Eq.\,(\ref{h1})
then gives 
\begin{equation} \label{s1}
S_1 \, = \, - \, 
e^{- p z} \, e^{- p' z'} \, \hat{h}({\bf p} + {\bf p}') \, \, .
\end{equation}
For example, for a sine modulation
with wavelength $2 \pi / k$ along, say, the $x$-axis, this implies 
that the incident wave vector component $p_x$ is scattered to 
$p_x' = p_x \pm k$ while the other components of ${\bf p}$
remain unchanged. 

A similar behavior occurs
for the contribution to $S$ at second order 
in $h$, which is given by
\begin{equation} \label{s2}
S_2 = - \, 
e^{- p z} \, e^{- p' z'}
\int \frac{d^D k}{(2 \pi)^D} \, |{\bf p} - {\bf k}| \,
\hat{h}({\bf k}) \, \hat{h}({\bf p} + {\bf p}' - {\bf k}) \, .
\end{equation}
For the sine modulation, this implies that $p_x$ is scattered by
$2 k$, $0$, $- 2 k$. 
In a scattering experiment with grazing incidence, the length 
scale perpendicular to the surface is set by the depth $b$
the evanescent wave penetrates the sample, giving rise to
diffuse scattering and thereby probing the critical correlations 
close to the surface \cite{DW84}. Since this diffuse scattering 
appears in addition to the contribution already present away
from criticality \cite{DH95}, it can in principle be separated 
out by tuning the temperature deviation $T - T_c$. 
We assume that $b$ is much larger than the 
height of the deformations.
In this case, the above expansion in the deformations results
in an expansion in powers of $h / b \ll 1$ for the 
elastic scattering cross section, which 
allows one to distinguish the corresponding contributions
via their intensities.

The second order results are particularly useful when dealing
with rough surfaces, where the quench averaged first order corrections
vanish. In particular, we shall assume that the rough surface is 
described by a height function $h({\bf x})$ with 
$\overline{ h(\bf x) } = 0$, and 
\begin{eqnarray}
\overline{ [h({\bf x}) - h({\bf y})]^2 } \, & = & \, 
\omega^{2 - 2 \zeta} \, |{\bf x} - {\bf y}|^2 \label{crossover}\\
& & \times \int \frac{d^D p}{(2 \pi)^D} \, 
e^{i {\bf p} \cdot ({\bf x} - {\bf y} ) } \, 
p^{-D + 2 - 2 \zeta} \, e^{-p \lambda} \, \, , \nonumber
\end{eqnarray}
where the overbar denotes averaging over different surface profiles.
While at large separations the above correlations grow as 
$|{\bf x} - {\bf y}|^{2 \zeta}$, we have also introduced a 
cutoff length $\lambda$ to regulate the behavior of the surface 
at short distances, and an overall amplitude length $\omega$. 

A characteristic feature of self-affine roughness is statistical
translational
invariance, since the rhs of Eq.\,(\ref{crossover}) depends
on the distance $|{\bf x} - {\bf y}|$ only. 
This implies that the averaged structure factor 
$\overline{S}$ is proportional to
$\delta^D({\bf p} + {\bf p}')$, and depends on 
$z$, $z'$, and $p$ = $|{\bf p}|$ only.
In order to maintain translational invariance, 
it is convenient to express the 
results for the correlation functions in terms of the local distance 
$\delta = z - h({\bf r}_{\parallel})$ from the surface rather than $z$
(see Fig.\,\ref{fig1}).
The two-point correlation function must now vanish as
$\delta$ or $\delta'$ go to zero. In terms of these coordinates,
the leading power law behavior of correlations is the same as for 
a flat surface, but the corresponding amplitude depends on 
the roughness and is modified by a factor of
$[1 - \alpha \, (\omega / \lambda)^{2(1-\zeta)}]$ as compared to
a flat surface, where $\alpha > 0$ is a number of order unity.
The subleading correction of order $\overline{h^2}$ decays with 
the separation $r = |\underline{r} - \underline{r}'|$ 
with an additional factor of $r^{-2(1-\zeta)}$ compared 
to the leading term.

For the interacting field theory, governed by Eq.\,(\ref{hamiltonian})
with $u_0 \neq 0$, standard perturbation theory can be applied to get
the correlation function
\widetext
\begin{equation} \label{loop2}
\langle \Phi_i(\underline{r}) \Phi_i(\underline{r}') \rangle \, = \,
G(\underline{r} ; \underline{r}') \, - \,
\frac{N + 2}{3} \, \frac{u_0}{2} \,
\int d^d R \, \, 
G(\underline{r} ; \underline{R}) \,
G(\underline{R} ; \underline{R}) \, G(\underline{R} ; \underline{r}\/')
\, \, + \, {\cal O}(u_0^2) \, \, .
\end{equation}
\narrowtext
\noindent
For a flat surface, the one-loop addition in $u_0$ can 
be regularized and renormalized by minimal subtraction of 
poles in $\varepsilon = 4 - d$, leading to logarithmic
contributions in the separation $r = |\underline{r} - \underline{r}'|$.
This perturbative result can then be improved by RG,
resulting in power laws in $r$
with corresponding surface critical exponents \cite{Bin83,Die86}.
For a self-affinely rough surface, 
the expansion of $G$ to second order in $h({\bf x})$ is substituted 
in the above equation, and the quench average is obtained using
Eq.\,(\ref{crossover}). The one-loop correction then 
gives rise to six diagrams of order $u_0 \overline{h^2}$. 
Similarly as for a flat surface,
the new diagrams in the quench average at order of 
$u_0 \overline{h^2}$ produce logarithmic contributions in $r$, 
which can again be recast as power laws. 
The final results for the two-point correlation function 
at criticality are summarized below.

Perpendicular correlations are obtained when $\underline{r}$ moves 
into the bulk while $\underline{r}'$ remains close to the surface, 
which implies $r = |\underline{r} - \underline{r}'| \to \infty$ while
$\delta'$ is fixed (see Fig.\,\ref{fig1}). In this case the 
correlations decay as
\begin{equation} \label{perp}
\overline{
\langle \Phi_i(\underline{r}) \Phi_i(\underline{r}') \rangle}
\, \sim \, \frac{1}{ r^{d-2+\eta_{\perp}}} \, + \,
\frac{a}{ r^{d-2+ \widetilde{\eta}_{\perp}}} \, \, ,
\end{equation}\\[-4mm]
where the first term corresponds to a flat surface with 
$\eta_{\perp} = 1 -  \frac{1}{2} \frac{n+2}{n+8} \varepsilon
\, + \, {\cal O}(\varepsilon^2)$. 
The second term describes the 
effect of self-affine roughness, with an amplitude $a$
depending on $\omega$, $\lambda$, and $\zeta$,
and the new universal exponent
\begin{equation} \label{etaperptilde}
\widetilde{\eta}_{\perp} \, = \, (2 - 2 \zeta) \, + \, 
1 \, - \, 2 \, \frac{n+2}{n+8} \, \varepsilon
\, \, + \, {\cal O}(\varepsilon^2) \, \, .
\end{equation}\\[-4mm]
Similarly, when both points remain close to the surface, 
i.e., both $\delta$ and $\delta'$ are fixed, correlations 
fall off as
\begin{equation} \label{para}
\overline{
\langle \Phi_i(\underline{r}) \Phi_i(\underline{r}') \rangle}
\, \sim \, \frac{1}{ r^{d-2+\eta_{\parallel}}} \, + \,
\frac{a'}{ r^{d-2+ \widetilde{\eta}_{\parallel}}} \, \, .
\end{equation}\\[-4mm]
In this case the flat surface is governed by 
$\eta_{\parallel} = 2 - \frac{n+2}{n+8} \, \varepsilon
+ {\cal O}(\varepsilon^2)$,
while self-affine roughness gives
\begin{equation} \label{etaparalleltilde}
\widetilde{\eta}_{\parallel}
\, = \, (2 - 2 \zeta) \, + \,
2 \, - \, 4 \, \frac{n+2}{n+8} \, \varepsilon \, \,
+ \, {\cal O}(\varepsilon^2) \, \, .
\end{equation}\\[-4mm]
The corrections due to roughness now decay with a slower
power as compared to the Gaussian case. Indeed, for a sufficiently
large roughness exponent $\zeta$, these corrections can even dominate 
the result for the flat surface. The borderline roughness exponent is
$\zeta_{\perp}^{\ast} = 1 - \frac{3}{4} \,
\frac{n+2}{n+8} \, \varepsilon \, + \, {\cal O}(\varepsilon^2)$
for perpendicular, and a different value of
$\zeta_{\parallel}^{\ast} \, = \, 1 \, - \, \frac{3}{2} \, 
\frac{n+2}{n+8} \, \varepsilon \, + \, {\cal O}(\varepsilon^2)$
for parallel correlations.
This is a surprising result from a naive point of view since,
due to $\zeta < 1$, on larger and larger length scales 
a self-affine rough surface looks more and more like a flat 
surface. Note that this effect becomes only visible beyond 
the Gaussian approximation, which corresponds to $\varepsilon = 0$.

\vspace*{15mm}

To test the generality of this result, we examined the correlations
for a $d = 2$ dimensional XY model below the Kosterlitz-Thouless
temperature \cite{KT78}. Correlations in the spin variables
$s(\underline{r}) = e^{i \theta(\underline{r})}$ 
decay with power laws in this case.
We found that the surface correlations fall off with the simple 
relative factor of $r^{-2(1-\zeta)}$. 
We attribute this to the 
Gaussian nature of the fluctuations in the phase angle 
$\theta(\underline{r})$, which are retained in the asymptotics 
of correlations for $s(\underline{r})$.

Thermodynamic quantities
and correlation functions can be obtained from derivatives of the 
free energy with respect to magnetic fields. To discuss
surface behavior, we introduce distinct fields $h_b$ and $h_s$
in the bulk and close to the surface, respectively. 
Assuming that our 
underlying assumption of the validity of an expansion in 
$h({\bf x})$ holds, the results for the two-point correlation
function are consistent with the following form for the scaling 
of the leading singular part of the surface free energy 
per projected area, 
\begin{eqnarray}
f_s^{(\text{sing})} \, = \, \xi^{-d+1\,}
& & \Big[ \, g_s(h_{b\,} \xi^{y_b}, h_{s\,} \xi^{y_s}) \label{ansatz}\\
& & \, \, + \, \xi^{- 2 (1-\zeta)} \, g_r(h_{b\,} \xi^{y_b}, 
h_{s\,} \xi^{\widetilde{y}_s}) \Big] \, \, , \nonumber
\end{eqnarray}
where $\xi \sim |T - T_c|^{- \nu}$
is the correlation length that diverges at the critical
point. The first term in square brackets
corresponds to a flat surface with $y_b$ and $y_s$
describing the relevance of 
bulk and surface fields, respectively \cite{Bin83,Die86}.
The second term gives the effect of surface roughness, with 
$\xi^{-2(1-\zeta)}$ reflecting the average increase in area.
However, to regain the results 
in Eqs.\,(\ref{perp}) - (\ref{etaparalleltilde})
we have to use a value of 
$\widetilde{y}_s = 1 + \frac{3 n}{2 (n+8)} 
\, \varepsilon \, + \, {\cal O}(\varepsilon^2)$
which is different from 
$y_s = 1 - \frac{3}{n+8} 
\, \varepsilon \, + \, {\cal O}(\varepsilon^2)$.
To motivate and justify this difference, we resort to
an analogy in which the rough surface is replaced with 
a collection of edges with a (possibly scale-dependent)
distribution of opening angles. Already for a single edge,
describing correlations requires a distinct
and angle-dependent value of $y_e$ for the magnetic field 
close to the edge \cite{Car83,HKSD99}. Similarly,
results obtained recently for correlations in the vicinity of a 
fractal surface with fractal dimension $d_f$ \cite{Dup98,Car99}
cannot be obtained using the value of $y_s$ for a flat surface
[with $\xi^{-d_f}$ replacing $\xi^{-d+1}$ in Eq.\,(\ref{ansatz})
and omitting the second term in square brackets].
Thus $\widetilde{y}_s$
can be regarded as inherently related 
to self-affine geometry. Interestingly, however, it does not 
depend on the roughness exponent $\zeta$, 
at least to order $\varepsilon$.

Future extensions of this research could focus on other
surface universality classes, the possibility of multifractal 
correlations, Casimir-type effects, and in particular exact 
results in two dimensions. However, our results derived here 
already allow for meaningful tests by simulations and experiments.

We thank J. Cardy, S. Dietrich, T. Emig, and R. Golestanian 
for helpful discussions and conversations.
This work was supported by the Deutsche Forschungs\-gemeinschaft
through grant HA3030/1-1 (AH) and the National Science 
Foundation through grant DMR-98-05833 (MK).

\end{multicols}\widetext

\end{document}